\documentclass{ws-ijbc}

\usepackage[normalem]{ulem}

\usepackage{enumitem}
\usepackage{multicol}
\usepackage{ws-rotating}     
\usepackage{graphicx}
\usepackage{epstopdf}
\usepackage{tabularx}
\usepackage[titlenumbered,ruled]{algorithm2e}
\usepackage{listings}
\usepackage{lmodern}
\usepackage[usenames,dvipsnames]{xcolor}

\usepackage{algorithmicx}
\usepackage{algpseudocode}
\usepackage{xcolor}

\usepackage[toc,page]{appendix}
\DeclareMathOperator{\sgn}{sgn}

\algrenewcommand\alglinenumber[1]{\tiny #1:}
\usepackage{mathtools}
\usepackage{amsmath}%
\usepackage{amsfonts}%
\usepackage{amssymb}%
\usepackage{graphicx}
\usepackage{color}
\usepackage{cases}
\def\HiLi{\leavevmode\rlap{\hbox to \hsize{\color{yellow!50}\leaders\hrule height .8\baselineskip depth .5ex\hfill}}}

\newlist{myenumi}{description}{10}
\setlist[myenumi]{labelindent=\parindent, leftmargin=*, label=(\roman*), align=left}
\setlist[myenumi]{leftmargin=0pt}

\definecolor{darkgreen}{rgb}{0.1, 0.5, 0.1}
\begin{document}

\markboth{M.-F. Danca, N.V. Kuznetsov}{Matlab code for Lyapunov exponents of fractional order systems}

\title{Matlab code for Lyapunov exponents of fractional order systems}

\author{MARIUS-F. DANCA}

\address{Romanian Institute od Science and Technology,\\
400487 Cluj-Napoca, Romania,\\
danca@rist.ro}

\author{NIKOLAY KUZNETSOV }
\address{Dept. of Applied Cybernetics, Saint-Petersburg State University, Russia \\and\\
Dept. of Mathematical Information Technology, University of Jyv\"{a}skyl\"{a}, Finland\\
nkuznetsov239@gmail.com}

\maketitle

\begin{history}
\received{(to be inserted by publisher)}
\end{history}

\begin{abstract}
In this paper the Benettin-Wolf algorithm to determine all Lyapunov exponents for a class of fractional-order systems modeled by Caputo\textsc{\char13}s derivative and the corresponding Matlab code are presented. First it is proved that the considered class of fractional-order systems admits the necessary variational system necessary to find the Lyapunov exponents. The underlying numerical method to solve the extended system of fractional order, composed of the initial value problem and the variational system, is the predictor-corrector Adams-Bashforth-Moulton for fractional differential equations. The Matlab program prints and plots the Lyapunov exponents as function of time. Also, the programs to obtain Lyapunov exponents as function of the bifurcation parameter and as function of the fractional order are described. The Matlab program for Lyapunov exponents is developed from an existing Matlab program for Lyapunov exponents of integer order. To decrease the computing time, a fast Matlab program which implements the Adams-Bashforth-Moulton method, is utilized. Four representative examples are considered.
\end{abstract}



\section{Introduction}
\begin{multicols}{2}

Despite a long history, the doubts that fractional-order (FO) derivatives have no clear geometrical interpretations (see e.g. \cite{pod}), was one of the several reasons that fractional calculus was not used in physics or engineering. However, during the last more than 10 years, fractional calculus starts to attract increasing attention. There are nowadays more and more works on FO systems and their related applications in physics, engineering, mathematics, finance, chemistry, and so on. For the theory on the existence, uniqueness, continuous dependence on parameters and asymptotic stability of solutions of FDEs with general nonlinearities see, for example, \cite{old,cap}, and \cite{existenta,kill,pod2}.

The Lyapunov exponents (LEs) measure the average rate of divergence
or convergence of orbits starting from nearby initial points.
Therefore, they can be used to analyze the stability of limits sets and
to check sensitive dependence on initial conditions,
that is, the presence of potential chaotic attractors.

On the other side, in \cite{ChaosBook} Cvitanovi\'{c} et al.
do not recommend the evaluation of the LEs and recommend to: ``Compute stability exponents and the associated covariant vectors instead. Cost less and gets you more insight. [...] we are doubtful of their utility as means of predicting any observables of physical significance''.
Moreover, we additionally note here, Perron's counterexample \cite{LeonovK-2007}
which shows actually that the use of LEs, obtained via the linearization procedure,
for the study of the behaviour of nonlinear system requires a rigorous justification.
However, determining LEs remains the subject of many works and
grown into a real software industry for modern nonlinear physics
(see, e.g.\cite{HeggerKS-1999,BarreiraP-2001,Skokos-2010,CzornikNN-2013,PikovskyP-2016,VallejoS-2017} and others).

Nowadays there are two widely used definitions\footnote{
Relying on the Oseledec ergodic theorem \cite{Oseledec-1968}
the above definitions often do not differ
(see, e.g. Eckmann \& Ruelle \cite[p.620,p.650]{lili},
Wolf~et~al.~\cite[p.286,p.290-291]{WolfSSV-1985},
and Abarbanel~et~al.~\cite[p.1363,p.1364]{AbarbanelBST-1993}),
however in general case, they may lead to different values
\cite{kuz1},\cite[p.289]{BylovVGN-1966},\cite[p.1083]{LeonovK-2007}.
}, of the LEs: via the exponential growth rates of norms of the fundamental matrix columns  \cite{Lyapunov-1892}
and via the exponential growth rates of the sigular values of fundamental matrix
\cite{Oseledec-1968}.
Corresponding approaches for the LEs computation and their difference
are discussed, e.g., in \cite{kuz1,KuznetsovAL-2016}.

Remark that in numerical experiments we can consider only finite time,
and, thus, the numerically computed values of LEs
can differ significantly from the limit values
(e.g. if the considered trajectory belongs to a \emph{transient chaotic set}),
and are often referred as \emph{finite-time LEs}.

Applying the statistical physics approach and assuming the ergodicity
(see, e.g. \cite{Oseledec-1968}),
the LEs for a given dynamical system are often estimated
by local LEs along a ``typical'' trajectory.
However, in numerical experiments,
the rigorous use of the ergodic theory is a challenging task
(see, e.g. \cite[p.118]{ChaosBook}).
If the LEs are the same for any trajectory,
then Frederickson~et~al. \cite[p.190]{FredericksonKYY-1983}
suggested to call them as \emph{absolute} ones and wrote that such
absolute values rarely exist.
For example, substantially different
values of the local LEs can be obtained along trajectories
on coexisting nonsymmetric attractors in the case of multistability\footnote{
While trivial attractors (stable equilibrium points)
can be easily found analytically or numerically,
the search of all periodic and chaotic attractors for a given system
is a challenging problem.
See, e.g. famous 16th Hilbert problem \cite{Hilbert-1901}
on the number of coexisting periodic attractors in two dimensional polynomial systems,
which was formulated in 1900 and is still unsolved,
and its generalization for multidimensional systems
with chaotic attractors \cite{LeonovK-2015-AMC}.
}
(see, e.g. such corresponding examples for the classical Lorenz system and Henon map
\cite{LeonovKKK-2016-CNSCS,KuznetsovLMPS-2018}).

In order to study the chaoticity of an attractor in numerical experiments,
one has to consider a grid of point covering the attractor and
compute corresponding  finite-time local LEs for a certain time.
Remark that while the time series obtained from a \emph{physical experiment}
are assumed to be reliable on the whole considered time interval,
the time series produced by the integration of
\emph{mathematical dynamical model}
can be reliable on a limited time interval only due to computational errors.

Taking into account the above discussion further
we consider finite-time local LEs
and their computation in Matlab by the analog of the Bennetin-Wolf algorithm.

\section{Benettin-Wolf Algorithm for LEs of FO }
The determination of LEs of a system of integer or fractional order with the Benettin-Wolf algorithm, requires the numerical integration of differential equations of integer or fractional order. Because the purpose of this paper is to present a Matlab code for LEs of systems of FO, in the next subsections, only the most important steps (such as the existence of variational equations of FO) used to implement the algorithm in Matlab language are presented (theoretical details can be found in the related references).

\subsection{Numerical integration of FDEs}

The autonomous FO systems considered in this paper are modeled by the following Initial Value Problem (IVP) with Caputo's derivative
\begin{equation}\label{unu}
\begin{array}{l}
D_*^qx=f(x),\\
x(0)=x_0,
\end{array}%
\end{equation}

\noindent for $t\in[0,T]$, $q\in(0,1)$, $f:\mathbb{R}^n\rightarrow \mathbb{R}^n$ and $D_*^q$, Caputo's differential operator of order $q$ with starting point $0$\footnote{Based on philosophical arguments rather than a mathematical point of view, some researchers questioned the appropriateness of using initial conditions of the classical form in the
Caputo derivative \cite{pil1}. However, it should be emphasized that, in practical (physical) problems, physically interpretable initial
conditions are necessary and Caputo\textsc{\char13}s derivative is a fully justified tool \cite{pil2}.}

    \begin{equation*}
    D_*^qx(t)=\frac{1}{\Gamma(1-q)}\int_0^t(t-\tau)^{-q}x'(\tau)d\tau,
    \end{equation*}
    with $\Gamma$ the known Euler function.

Properties of the Caputo's differential operator, $D_*^q$, are discussed in \cite{pod2,gore}.

Under Lipschitz continuity of the function $f$, the IVP \eqref{unu} admits a unique solution \cite{kai}.

\begin{remark}\label{remarcus}
\vspace{-3mm}
\begin{itemize}\setlength{\itemindent}{.3in}
\item[\emph{i})]
In the case of an integer-order dynamical system, denoting the solution of the underlying IVP as $x(t,x_0)$,
one has
$x_s\circ x_t=x_{t+s}$
(see e.g. \cite{Z}). Due to the memory dependence of the derivatives, this does not hold in the case of systems modeled by FDEs. However, motivated by the numerical character of this paper, the definition of integer order dynamical systems which states that
if the underlying IVP admits unique solutions existing on infinite time interval,
the problem defines a dynamical system (see \cite[Definition 2.1.2]{stu}) is adopted.
\item[\emph{ii})] Even fractional-order dynamics describe a real object more accurately than classical integer-order dynamics, systems modeled by the IVP \eqref{unu} cannot have any non-constant periodic solution (see e.g. \cite{tava}). However, a solution may be asymptotically periodic \cite{dada0}. These trajectories are called \emph{numerically periodic}, in the sense that the trajectory, from numerical point of view, can be an extremely-near periodic with respect to, e.g., Euclidean norm \cite{dada0}. A numerically periodic trajectory refers to as a closed trajectory in the phase space in the sense that the closing error is within a given bound of $1E-n$, with $n$ being a sufficiently large positive integer.
\end{itemize}
\end{remark}

The numerical integrations required by the LEs algorithm for FO systems are performed in this paper with the predictor-corrector Adams-Bashforth-Moulton (ABM) method for FDEs,
proposed Diethelm et al. \cite{kai} which is constructed for the fully general set of equations without any special assumptions, being easy to implement in any language.

Let us next assume that we are working on a uniform grid $\{t_{j}=jh:n=0,1,...,N+1\}$ with some integer $N$ with the step size
$h$ and the case of $q\in (0,1)$. Then, the predictor form, $x^{P}$, at the point $t_{j+1}$, is the
fractional variant of the Adams-Bashforth method
\begin{equation*}
x^{P}\left( t_{n+1}\right) =x_{0}+\frac{1}{\Gamma \left( q\right) }%
\sum\limits_{j=0}^{n}b_{j,n+1}f\left( x\left( t_{j}\right) \right) ,
\end{equation*}
\noindent while the corrector formula (the fractional variant of the
one-step implicit Adams Moulton method) reads
\begin{eqnarray*}
x\left( t_{n+1}\right) =x_{0}+\frac{h^{q}}{\Gamma (q+2)}f\left(
x^{P}\left( t_{n+1}\right) \right)\\
+\frac{h^{q}}{\Gamma (q+2)}\sum\limits_{j=0}^{n}a_{j,n+1}f\left( x\left(
t_{j}\right) \right) ,
\end{eqnarray*}
\noindent where $a$ and $b$ are
the corrector and predictor weights respectively given by the
following formula

\begin{equation*}
a_{j,n+1}=\left\{
\begin{array}{ll}
n^{q+1}-\left( n-q\right) \left( n+1\right) ^{q} & \text{if }j=0, \\
\left( n-j+2\right) ^{q+1}+\left( n-j\right) ^{q+1}\\ -2\left(
n-j+1\right)
^{q+1} & \text{if }1\leq j\leq n, \\
1 & \text{if}~j=n+1,%
\end{array}%
\right.  \label{a}
\end{equation*}
\noindent and
\begin{equation*}\label{b}
b_{j,n+1}=\frac{h^{q}}{q}\left( \left( n+1-j\right) ^{q}-\left( n-j\right)
^{q}\right) .
\end{equation*}

\begin{table}[h]
\begin{center}
\begin{tabular}{lll}
\hline

for~$k:=1~$to$~N~$do \\
$\ \ \ \ a[k]:=k^{q}-\left( k-1\right) ^{q}$ \\
$\ \ \ \ b[k]:=\left( k+1\right) ^{q+1}-2k^{k+1}+\left( k-1\right) ^{q+1}$
\\
end \\
\hline

\end{tabular}%
\end{center}
\end{table}

The predictor-corrector ABM method has an error which is roughly proportional to $h^2$. Thus, to obtain an error of, e.g., $1.0E-6$, a step size close to $h=1.0E - 3$ should be considered.

\hspace*{6mm} Because due to the long memory processes, the utilized ABM method as described in \cite{kai}, is time consuming. Therefore, a fast optimized ABM method, \texttt{FDE12.m} \cite{roberto} is used.

\texttt{FDE12.m} is called by the following command line:
\vspace{3mm}
\begin{lstlisting}
[t,x] = FDE12(q,fcn,t0,tf,x0,h);
\end{lstlisting}
\vspace{3mm}
\noindent where \texttt{q} represents the commensurate fractional-order, $\texttt{fcn.m}$ is the file with the function to be integrated, \texttt{t0} and \texttt{tf} define the time span, \texttt{x0} are the initial conditions, and \texttt{h} represents the integration step-size.

For example, consider the integration of the FO Rabinovich-Fabrikant (RF) system \cite{dada2}, which has the following Matlab function, $\texttt{RF.m}$
\vspace{3mm}
\begin{lstlisting}
function dx = RF(t,x)
 dx=[x(2)*(x(3)-1+x(1)*x(1))+0.1*x(1);
    x(1)*(3*x(3)+1-x(1)*x(1))+0.1*x(2);
    -2*x(3)*(p+x(1)*x(2))];
\end{lstlisting}
\vspace{3mm}
with the bifurcation parameter $p=0.98$.

With the following parameters, one obtains the chaotic attractor in Fig. \ref{fig1} (a):
\vspace{3mm}
\begin{lstlisting}
[t,x]=FDE12(0.999,@RF,0,1500,[0.1;0.1;0.1],...
0.01);
\end{lstlisting}
\vspace{3mm}
To note that \texttt{FDE12.m} requires the entry, \texttt{x0}, as a column vector \texttt{[x10,x20,x30]'} or, similarly, \texttt{[x10;x20;x30]}. Also, the returned exit, \texttt{x}, is a column vector.

\subsection{Algorithm for LEs of the FO system \eqref{unu}}

\begin{notation}
Hereafter the finite-time local LEs of FO
are called LEs.
\end{notation}

The existence of the variational equations necessary to determine LEs is ensured by the following Theorem \cite{exist}

\begin{theorem}
System \eqref{unu} has the following variational equations which define the LEs
\begin{equation}\label{doi}
\begin{array}{l}
D_*^q \Phi(t)=D_xf(x)\Phi(t),\\
\Phi(0)=I,
\end{array}
\end{equation}
\noindent where $\Phi$ is the matrix solution of the system \eqref{unu}, $D_x$ is the Jacobian of $f$ and $I$ is the identity matrix.
\end{theorem}

Further we assume that for the matrix $\Phi(t)$ the cocycle property takes place.

\begin{figure*}[ht]
\centering
\begin{minipage}[t]{0.7\textwidth}
\SetInd{2ex}{1ex}
\begin{algorithm*}[H]\label{LEalgo}
\caption{Algorithm for LEs of FO system \eqref{unu}}
\SetKwInOut{Input}{Input}
    \Input{} 
-$ne$\Comment{{number of equations }}
\\-$x\_start$ \Comment{{$n_e$ initial conditions of \eqref{unu}}}
\\-$t\_start$, $t\_end$\Comment{{time span}}
\\-$h\_norm$\Comment{{Normalization step-size}}
\\$n\_it\gets$ $(t\_end-t\_start)/h\_norm$ \Comment{{iterations number}}
\\  \For{$i\gets$ $ne+1$~\KwTo $ne(ne+1)$}
{$x(i)=1.0$\Comment{{initial conditions of \eqref{doi}}}
}
 {$t\gets t\_start$}
\\\For{$i\gets$ $1$~\KwTo $n\_it$}
{
\HiLi$x\gets$ \text{integration of FO systems \eqref{unu}-\eqref{doi}}
\\{$t\gets t+h\_norm$}
\\{$zn(1),...,zn(ne)\gets$ \text{\emph{Gram-Schmidt procedure}}}
\\{$s(1)\gets 0$}
\\\For{$k\gets$ $1$~\KwTo $ne$}
{
$s(k)\gets$ $s(k)+log(zn(k))$\Comment{{vector magnitudes}}
\\$LE(k)\gets$ $s(k)/(t-t\_start)$\Comment{{\texttt{LEs}}}
 }
}
\textbf{Output}:LE
\end{algorithm*}
\end{minipage}
\end{figure*}

Therefore, the algorithm to determine the LEs of the system \eqref{unu} becomes similar to the case of integer order.

Lyapunov exponents measure the exponential growth, or decay, of infinitesimal phase-space perturbations of a chaotic dynamical system.

The algorithm for numerical evaluation of LEs utilized in this paper, has been proposed in the seminal works of Benettin et al. \cite{bene2} (see also \cite{shima}), one of the the first work to propose a Gram-Schmidt orthogonalization procedure to compute LEs for continuous systems of integer order, as described in \cite{lili}), and by Wolf et al. \cite{lyappp} (see also \cite{eci}).

The algorithm to find all LEs, described as a Fortran code by Wolf et al. \cite{lyappp}, and also as a Basic code in \cite{baker}, solves the equations of motion under perturbations and periodic orthonormalization.

To note that the accuracy and reliability of numerically determined LEs depend on initial conditions, on the selection of the perturbations, performances of the utilized integration numerical method and also on orthonormalization step size. A long-time numerical calculation of the
leading Lyapunov exponent requires rescaling the distance between nearby trajectories, in order to keep the separation within the linearized flow range. To avoid overflow, one calculates the divergence of nearby trajectories for finite timesteps and renormalizes to unity after a finite number of steps (Gram-Schmidt procedure
\cite{lili,christ}).


Therefore, the main steps to determine numerically the LEs are: numerical integration of the FO system \eqref{unu} together with the variational system \eqref{doi} (i.e. the extended system), Gram-Schmidt procedure
and picking up the exponents during the renormalization procedure, the LEs being determined as the average of the logarithm of the stretching factor of each perturbation, steps presented in Algorithm \ref{LEalgo}.

\section{The Matlab code for LEs}
Consider the following general assumptions:

\begin{itemize}[leftmargin=.2in]
\item The considered systems, modeled by the IVP \eqref{unu}, are autonomous;

\item The system \eqref{unu} is of commensurate order: $q_1=q_2=...=q_{ne}=q$;\footnote{
    The incommensurate case can be treated similarly, the only difference referring to the utilized numerical method for FDEs.}

\item In the case of chaotic behavior, the fractional-order has been chosen close to $1$, such that chaos is significant.

\item The right hand side of system \eqref{unu}, $f$, is smooth enough.

\item Because of the space restrictions, only the main programs code are presented, and indications on how to write the other ones.
\end{itemize}

A simple way to build in Matlab language the algorithm for FO systems, the program \texttt{FO\_Lyapunov.m} (Appendix \ref{a}), was to modify either some existing program, e.g., the program \texttt{lyapunov.m} \cite{govo}, which is a Matlab variant of the original LEs algorithm proposed in \cite{bene2} or \cite{lyappp} or, similarly, to translate in Matlab the BASIC program \cite{baker}, a close variant of the original algorithm) or, also, the Fortran code proposed by Wolf et all in \cite{lyappp}, and modify it for FDEs.

The program, called \texttt{FO\_Lyapunov.m}, is launched with the following command line:
\vspace{3mm}
\begin{lstlisting}
[t,LE]=FO_Lyapunov(ne,@ext_fcn,t_start,h_norm,...
t_end,x_start,h,q,out);
\end{lstlisting}
\vspace{3mm}
where \texttt{ne} represents the equations (and state variables) number, \texttt{ext}$\_$\texttt{fcn.m} the function containing the extended system \eqref{unu}-\eqref{doi}, \texttt{t}$\_$\texttt{start} and \texttt{t}$\_$\texttt{end} the time span, \texttt{h}$\_$\texttt{norm} the normalization step, \texttt{x}$\_$\texttt{start} the initial condition (as column vector), \texttt{h} the step size of FDE12.m, and \texttt{out} indicates the steps number when intermediate values of time and LEs are printed (for \texttt{out=0}, no intermediate results will be printed out).

As shown in the algorithm for LEs (Algorithm \ref{LEalgo}), it is necessary to solve the extended system \eqref{unu}-\eqref{doi} of FO (yellow line) which is given in the function \texttt{ext\_fcn.m}. In this file, beside the right hand side function $\texttt{f}$ of the system \eqref{unu}, the Jacobi matrix $\texttt{J}$ should be included (see Appendix \ref{b} where the function for the RF system is presented).

The program plots the time evolutions of the LEs.

\section{Numerical tests}

Beyond numerical artifacts that might occur when numerically integrating a system of ODEs of integer order, notions such as ``shadowing time'' and ``maximally effective computational time'' reveal that it is possible to have reliable numerical simulations only on a relative finite-time interval (see, e.g., \cite{sara,sara2}). The case of FO systems is even more delicate. In this paper we have considered generally $t\in[0,300]$ and, for the Lorenz system, $t\in[0,500]$.

\begin{figure*}
\begin{center}
\includegraphics[scale=0.63] {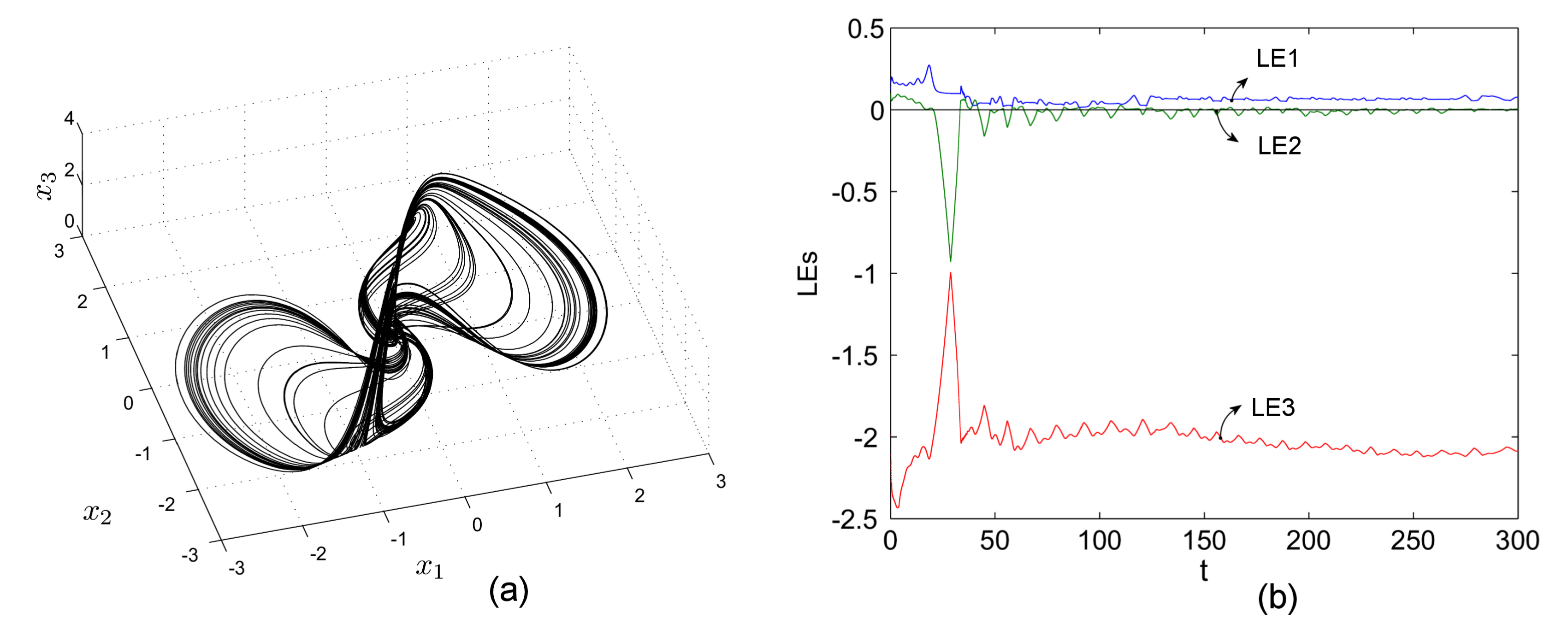}
\caption{(a) A chaotic attractor of the RF system of FO, for $q=0.999$. (b) Dynamics of the LEs.}
\label{fig1}       
\end{center}
\end{figure*}

\begin{figure*}
\begin{center}
\includegraphics[scale=0.7] {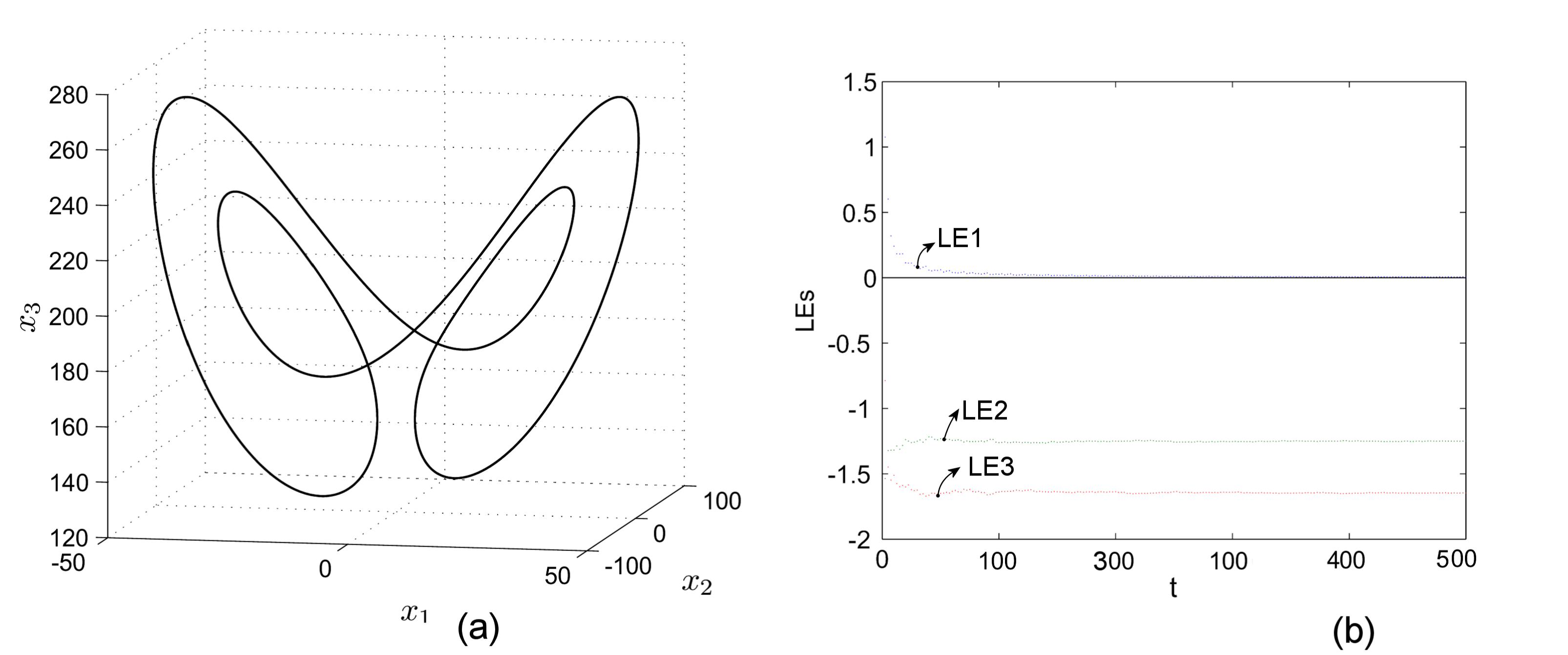}
\caption{(a) An apparently stable cycle of the generalized Lorenz system of FO, for $q=0.985$. (b) Dynamics of the LEs.}
\label{fig2}       
\end{center}
\end{figure*}

\begin{figure*}
\begin{center}
\includegraphics[scale=0.64] {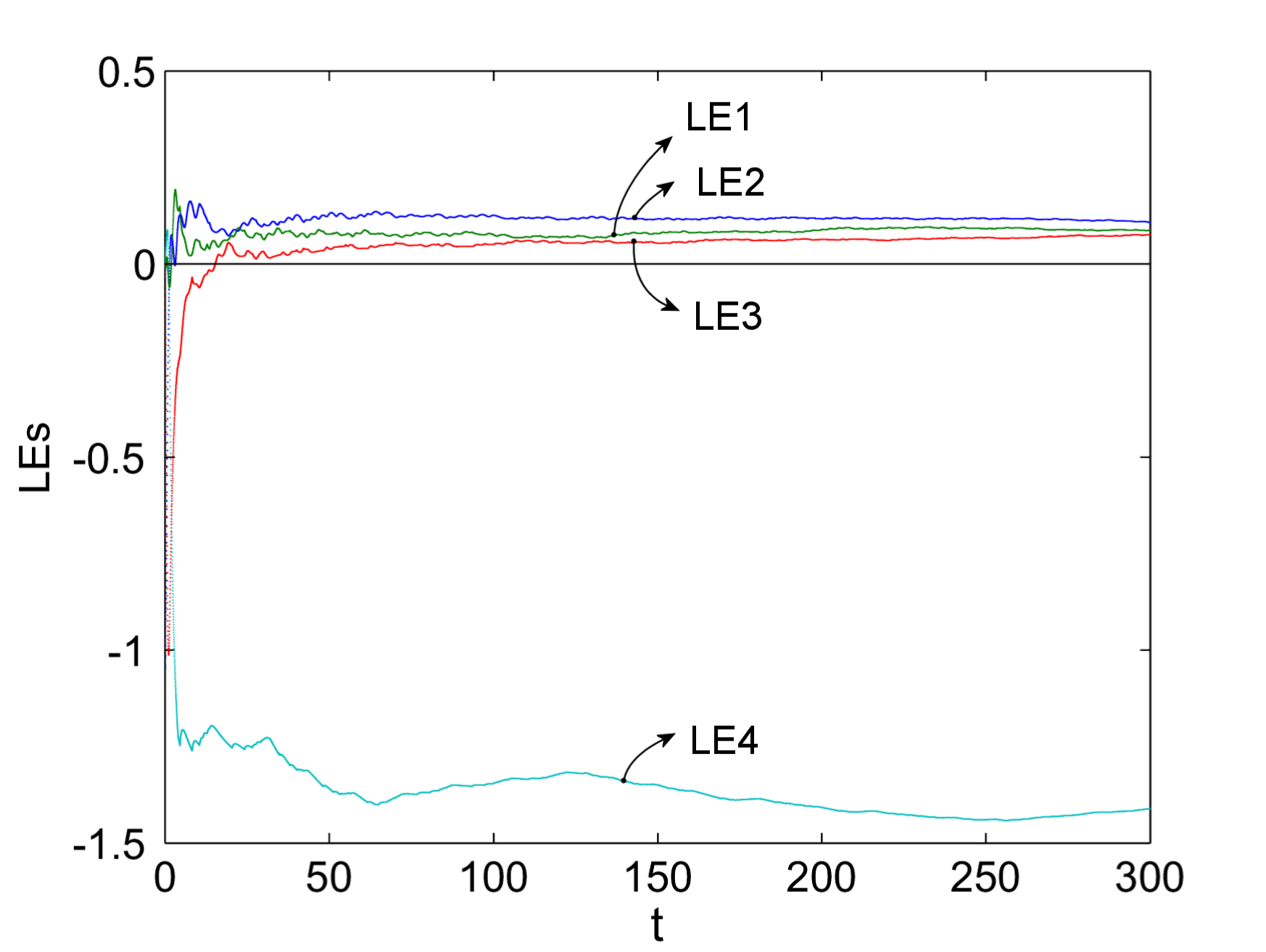}
\caption{Dynamics of the LEs of the 4-dimensional system of FO \eqref{sp}.}
\label{fig3}       
\end{center}
\end{figure*}

\begin{itemize}[leftmargin=.2in]
\item[1.]
Let consider the function for the RF system, $\texttt{LE\_RF.m}$, which include the extended system \eqref{unu}-\eqref{doi} (Appendix \ref{b}).
Because \texttt{ne=3}, beside the \texttt{3} variables \texttt{x(1),x(2),x(3)} required by the numerical solution of the original system \eqref{unu}, the matrix solution of the system \eqref{doi} requires other more \texttt{ne}$\times$\texttt{ne=9} variables from the total of \texttt{ne(ne+1)=12} variables: \texttt{x(1:12)}, \texttt{f=zeros(size(x))=zeros(12)}, where \texttt{f} loads the first \texttt{ne=3} righthand side expressions of system \eqref{unu}, and the \texttt{ne$\times$ne=9} righthand side expressions of the variational system \eqref{doi}.

For example, for the RF system, with the following command line:
\vspace{3mm}
\begin{lstlisting}
[t,LE]=FO_Lyapunov(3,@LE_RF,0,0.02,300,...
[0.1;0.1;0.1],0.005,0.999,1000);
\end{lstlisting}
\vspace{3mm}
one obtains the intermediary results printed every \texttt{out=1000} \texttt{h}$\_$\texttt{norm} steps, presented in Table \ref{tabis} (see also Fig. \ref{fig1} (b) where the dynamics of the LEs are drawn).
\vspace{3mm}
\lstset{%
    caption=Caption Text,
    frame=tb
  }
\begin{lstlisting}[label=tabis]
     10.00     0.1611     0.0660    -2.1614
     20.00     0.1923     0.0069    -2.0503
     30.00     0.0984    -0.7397    -1.1817
     40.00     0.0248     0.0542    -1.9761
     50.00     0.0440    -0.0168    -1.9867
     60.00     0.0697    -0.0006    -2.0701
     70.00     0.0354    -0.0116    -2.0167
     80.00     0.0331    -0.0618    -1.9360
     90.00     0.0201     0.0125    -1.9754
    100.00     0.0429     0.0112    -1.9794
    110.00     0.0337     0.0252    -1.9698
    120.00     0.0262    -0.0213    -1.9038
    130.00     0.0624    -0.0043    -1.9537
    140.00     0.0660    -0.0029    -1.9811
    150.00     0.0645    -0.0010    -2.0008
    160.00     0.0743     0.0018    -2.0304
    170.00     0.0710    -0.0009    -2.0394
    180.00     0.0583     0.0139    -2.0548
    190.00     0.0678     0.0042    -2.0665
    200.00     0.0662    -0.0217    -2.0498
    210.00     0.0628    -0.0155    -2.0622
    220.00     0.0602    -0.0125    -2.0715
    230.00     0.0570    -0.0222    -2.0666
    240.00     0.0643    -0.0222    -2.0813
    250.00     0.0639    -0.0079    -2.1020
    260.00     0.0623     0.0045    -2.1121
    270.00     0.0630     0.0006    -2.0987
    280.00     0.0551    -0.0036    -2.0771
    290.00     0.0761     0.0009    -2.0937
   <@  \textcolor{blue}{300.00}@>   <@  \textcolor{blue}{0.0749}@>   <@  \textcolor{blue}{0.0018} @> <@  \textcolor{blue}{-2.0850}  @>
\end{lstlisting}
Table 1. For $t\in[0,300]$, the RF system has \texttt{LE = (0.0749, 0.0018, -2.0850)} (last line, blue).
\vspace{3mm}
\lstset{%
    frame=single
  }

\item[2.]
If one considers the Lorenz system
\begin{equation}\label{sp}
\begin{array}{l}
D_*^{q}x_1=\sigma(x(2)-x(1)),\\
D_*^{q}x_2=-x(1)x(3)+px(1)-x(2)\\
D_*^{q}x_3=x(1)x(2)-\beta x(3);
\end{array}%
\end{equation}

with $q=0.985$ and the standard parameters $\sigma=10$, $\beta=8/3$ and the bifurcation parameter $p=200$, after some neglected transients, one obtains an apparently stable cycle (see Fig. \ref{fig2} (a) and Remark \ref{remarcus} (\emph{ii})).

To obtain the LEs one write the following command line:
\vspace{3mm}
\begin{lstlisting}
[t,LE]=FO_Lyapunov(3,@LE_Lorenz,0,5,500,...
[0.1;0.1;0.1],0.001,0.985,10);
\end{lstlisting}
\vspace{3mm}
which gives \texttt{LE=(-0.0026, -0.0870, -1.6225)}. The function \texttt{LE}$\_$\texttt{Lorenz.m} can be obtained similarly with \texttt{LE}$\_$\texttt{RF.m}.
The time evolution of the LEs is presented in Table 2 (see also Fig. \ref{fig2} (b)).
\lstset{%
    caption=Caption Text,
    frame=tb
  }
\begin{lstlisting}[label=tabu]
     50.00     0.1759    -0.1591    -1.5683
    100.00     0.0611    -0.1108    -1.6050
    150.00     0.0346    -0.0927    -1.6300
    200.00     0.0215    -0.0877    -1.6288
    250.00     0.0135    -0.0866    -1.6269
    300.00     0.0082    -0.0865    -1.6255
    350.00     0.0043    -0.0866    -1.6244
    400.00     0.0014    -0.0867    -1.6236
    450.00    -0.0008    -0.0869    -1.6230
   <@  \textcolor{blue}{500.00}@>  <@  \textcolor{blue}{-0.0026}@>  <@  \textcolor{blue}{-0.0870} @> <@  \textcolor{blue}{-1.6225}  @>
\end{lstlisting}
Table 2. Time evolution of the LEs for the Lorenz system. \texttt{LE=(-0.0026, -0.0870, -1.6225).}
\vspace{3mm}
\lstset{%
    frame=single
  }

\item[3.]

Consider next the non-smooth 4-dimensional system \cite{dada0}
\begin{equation}\label{sp}
\begin{array}{cl}
D_*^{q}x_1= & -x_{1}+x_{2}, \\
D_*^{q}x_2= & -x_{3}\sgn(x_1)+x_{4}, \\
D_*^{q}x_3= & |x_{1}|-a,\\
D_*^qx_4=&-bx_2,
\end{array}%
\end{equation}
with $a=1$, $b=0.5$ and $q=0.98$.

With the command line:
\vspace{3mm}
\begin{lstlisting}
[t,LE]=FO_Lyapunov(4,@LE_4d,0,0.02,300,...
[0.1;0.1;0.1;0.1],0.005,0.98,1000);
\end{lstlisting}
\vspace{3mm}
one obtains \texttt{LE=(0.1262, 0.0846, 0.0778, -1.5244)} (Fig. \ref{fig3}).

For this example, \texttt{ne=4}, the number of the variables is $20$ and, therefore, compared with systems with \texttt{ne=3}, the function file, \texttt{LE}$\_$\texttt{4d.m}, must be modified accordingly (compare the red line in \texttt{LE}$\_$\texttt{RF.m}).
Also, in order to obtain the printed intermediated values of LEs, in the \texttt{FO}$\_$\texttt{Lyapunov.m}, the \texttt{fprintf} command, must be modified by adding one supplementary specifier \texttt{\%10.3f}.

Because from the four LEs, the system admits three positive LEs, it can be considered as hyperchaotic (see \cite{dada} for a discussion about the number of positive LEs of hyperchaotic systems).

\begin{remark}
Note that because the system is not smooth and also discontinuous, the correct numerical integration required for this system cannot be done without a previous smooth approximation  \cite{dada0}. Thus, like all numerical methods for FDEs which are designed for continuous dynamical systems, the integrator \texttt{FDE12} cannot be utilized in this case without the mentioned approximation. Also, without a smooth approximation, the Jacobian matrix, utilized in the Benettin-Wolf algorithm, cannot be determined.
\end{remark}

\item[4.]

LEs can be also plotted as function of the bifurcation parameter $p$ for $p\in[p_{min},p_{max}]$ (program \texttt{run}$\_$\texttt{Lyapunov}$\_$\texttt{p.m}, Appendix \ref{c}).
The program uses a slightly modified variant of \texttt{FO}$\_$\texttt{Lyapunov.m}, which is called \texttt{FO}$\_$\texttt{Lyapunov}$\_$\text{p.m}. To obtain \texttt{FO}$\_$\texttt{Lyapunov}$\_$\text{p.m} from \texttt{FO}$\_$\texttt{Lyapunov.m}, the following modifications have to be done:

\begin{itemize}\setlength{\itemindent}{.2in}
\item[(a)] The header line of the code \texttt{FO}$\_$\texttt{Lyapunov.m} is replaced with the following line (to note that the output \texttt{t} and the input \texttt{out} are no more necessary)
\vspace{3mm}
\begin{lstlisting}
function LE=FO_Lyapunov_p(ne,ext_fcn_p,...
t_start,h_norm,t_end,x_start,h,q,p);
\end{lstlisting}
\vspace{3mm}
\item[(b)] the printing and plotting lines \texttt{(*)} (red color) are deleted;
\item[(c)] the rest of the lines remain the same;
\end{itemize}

The m-file containing the extended system, \texttt{ext}$\_$\texttt{fcn}$\_$\texttt{p.m}, for the RF system, is presented in Appendix \ref{d}. Passing the parameter $p$ between these codes, can be easily realized due of the facilities of the program FDE12 (see FDE12).

For example, for the RF system, for $p\in[1.1,1.3]$, for $1000$ values of $p$ ($n=1000$), with the following command
\vspace{3mm}
\begin{lstlisting}
run_FO_Lyapunov_p(3,@LE_RF_p,0,0.02,200,...
[0.1;0.1;0.1],0.002,0.998,1.1,1.3,800)
\end{lstlisting}
\vspace{3mm}
one obtains the evolution of the LEs drawn in Fig. \ref{fig4} (a).

\item[5.]

LEs can be plotted also as function of the fractional order \texttt{q} (program \texttt{run}$\_$\texttt{Lyapunov}$\_$\texttt{q.m} in Appendix \ref{e}).
The used program \texttt{FO}$\_$\texttt{Lyapunov}$\_$\texttt{q}, is obtained from \texttt{FO}$\_$\texttt{Lyapunov} with the following modifications:
\begin{itemize}\setlength{\itemindent}{.2in}
\item[(a)] The header line of the code is replaced with the following line (to note that the input \texttt{out} is no more necessary);
\vspace{3mm}
\begin{lstlisting}
function [t,LE]=FO_Lyapunov_q(ne,ext_fcn,...t_start,h_norm,t_end,x_start,h,q);
\end{lstlisting}
\vspace{3mm}
\item[(b)] the printing and plotting lines \texttt{(*)} (red color) are deleted;
\item[(c)] the rest of the lines remain the same;
\end{itemize}

The function containing the extended system, \texttt{ext}$\_$\texttt{fcn.m}, does not require any modification.

For example, for the RF system, with the command
\vspace{3mm}
\begin{lstlisting}
run_FO_Lypaunov_q(3,@LE_RF,0,0.05,150,...
[0.1;0.1;0.1],0.002,0.9,1,800)
\end{lstlisting}
\vspace{3mm}
one obtains the LEs plotted in Fig. \ref{fig4} (b).

\begin{remark}
The presented programs can be optimized especially in the cases of \texttt{FO}$\_$\texttt{Lyapunov}$\_$\texttt{p.m} and \texttt{FO}$\_$\texttt{Lyapunov}$\_$\texttt{q.m}. A simple improvement was to use \texttt{while} loop (instead \texttt{for}) which for these two programs reduce substantially the computational time. However, every parameter $p$ (or order $q$) step, several constant parameters are shared between programs, fact which slows significantly the programs. Therefore, supplementary optimization should be done.
\end{remark}

\item[6.]
Another potential application of the proposed LEs algorithm, is to represent the LEs as function of two variables: the order $q$ and the bifurcation parameter $p$.

Let the Chen system
\[
\begin{array}{l}
D_*^q x_1=a(x_2-x_1),\\
D_*^q x_2=(p-a)x_1-x_1x_3+cx_2,\\
D_*^q x_3=x_1x_2-bx_3,
\end{array}
\]

with parameters $a=35$, $b=3$ and $q$ and $p$ variables. By considering $S_i:=LE(q,c)$, for $i=1,2,3$, for $q\in[0.9,1]$ and $p\in[20,30]$, the obtained surfaces are plotted in Fig. \ref{fig5} (see \cite{dada} where the algorithm to obtain LEs as surfaces is described). One can see that there exists a unique positive LE (red surface, $S_1$) for all values of the considered parameter $c$ but only for some fractional order values $q$, for which $S_1$ is situated over the horizontal plane LE=0, where LEs are zero. Also, one can see that $S_2$, which is almost identic to $S_1$ when the underlying LE1,2, are negative, becomes zero for a relative large values of $q$ and $p$, when $S_2$ separates from $S_1$ and identify (with the underlying numerical error) with the plane LE=0.
\end{itemize}

\section{Conclusion and discussion}
Starting from an existing variant for integer-order system, in this paper, based on the Benettin-Wolf algorithm, we proposed a Matlab program to determine numerically
finite-time local LEs for FO systems.

Due to inherent numerical errors, the algorithm should be utilized with precaution.
As known, among the numerical errors, the results depends strongly on the initial conditions,
time integration interval and, especially, on the renormalization step size \texttt{h}$\_$\texttt{norm}.

The relative large numerical errors of the Benettin-Wolf algorithm, which for the considered examples was of order of $1.E-2$, can be observed in the cases when one knows that the system presents a numerically periodic cycle (see Remark \ref{remarcus} (\emph{ii})), when the maximum LE should be zero. For example, the Lorenz system, which for $p=200$ presents such apparently stable cycle
(see Fig. \ref{fig2} (a)), has the maximum LE zero only with two precise decimals. Actually, in general, three precise decimals for zero LEs are extremely rare. Note that this zero value, appears only for \texttt{h}$\_$\texttt{norm=5} and a smaller integration step size, $h=0.001$ and only for a larger time interval \texttt{[0,500]}. Therefore, if no special improvements of numerical integration are implemented, with Benettin-Wolf algorithm, for integer but also for fractional order systems, two or three decimals are the maximal expected number of decimals.

One of the most important algorithm variable, is \texttt{h}$\_$\texttt{norm}, which determines the normalization moments influences the results.
This dependence can be also deduced from the example of the RF system. With \texttt{h=0.005} and \texttt{h}$\_$\texttt{norm=0.005} one obtains \texttt{LEs=(0.0631, 0.0031, -2.0774)}, while with the same stapsize, \texttt{h=0.005}, but \texttt{h}$\_$\texttt{norm=0.5}, \texttt{LEs=(0.0643, 0.0026, -1.8254)}.

Since on our best knowledge, there is not criterion to choose precisely \texttt{h}$\_$\texttt{norm}, the recommended way would be to realize several tests to choose the value for which slightly modifications does not change significantly the results.

In the case of fixed step-size integration numerical methods, like the considered \texttt{FDE12}, \texttt{h}$\_$\texttt{norm} can be chosen as depending on the step size \texttt{h}. Therefore, as Fig. \ref{fig6} shows, \texttt{h}$\_$\texttt{norm} could be multiple of integration step size \texttt{h} ($\varepsilon$ represents the perturbation between nearly two trajectories $x$ and $\bar{x}$).

Regarding the Gram-Schmidt procedure, in order to speed up the code, one can use the QR decomposition based on
  the Householder transformation: {\ttfamily [Q, R] = qr(A)}.
  To have matrix $R$ with positive diagonal elements, one can additionally use
  {\ttfamily Q = Q*diag(sign(diag(R))); R = R*diag(sign(diag(R)))}.
  See also \cite{RamasubramanianS-2000}.

The integration step-size of \texttt{FDE12}, $h$, plays an important role. As the documentation of the program specifies, there exists the possibility to increase the performances of the numerical integration, but in time integration detriment.

\end{multicols}

\section*{Acknowledgements} \label{sec:acknowledgement}
The work is done within Russian Science Foundation project 14-21-00041.

\pagebreak

\begin{figure*}
\begin{center}
\includegraphics[scale=1.5] {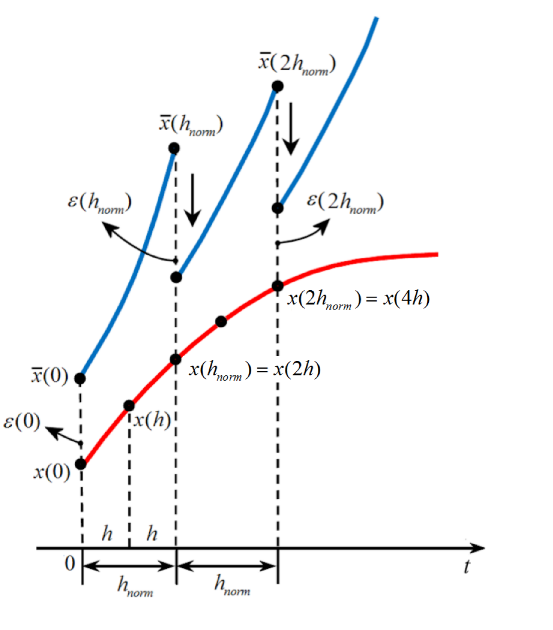}
\caption{Perturbation and rescaling of a nearby trajectory, after every $h_{norm}$ steps, considered as multiple of $h$ (here $h_{norm}=2h$, sketch).}
\label{fig4}       
\end{center}
\end{figure*}

\begin{figure*}
\begin{center}
\includegraphics[scale=0.6] {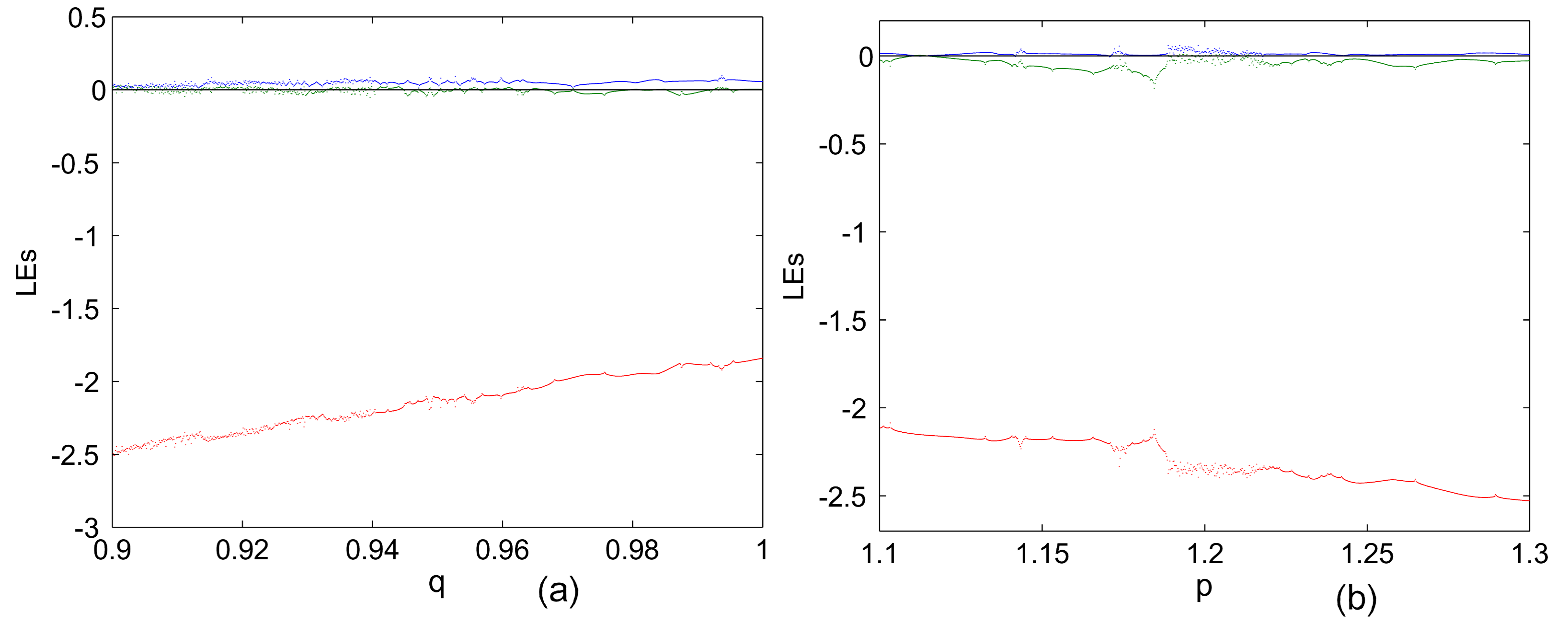}
\caption{LEs of RF system. (a) LEs as function of $q$ for $q\in[0.9,1]$. (b) LEs as function of $p$ for $p\in[1.1,1.3]$.}
\label{fig5}       
\end{center}
\end{figure*}

\begin{figure*}
\begin{center}
\includegraphics[scale=0.6] {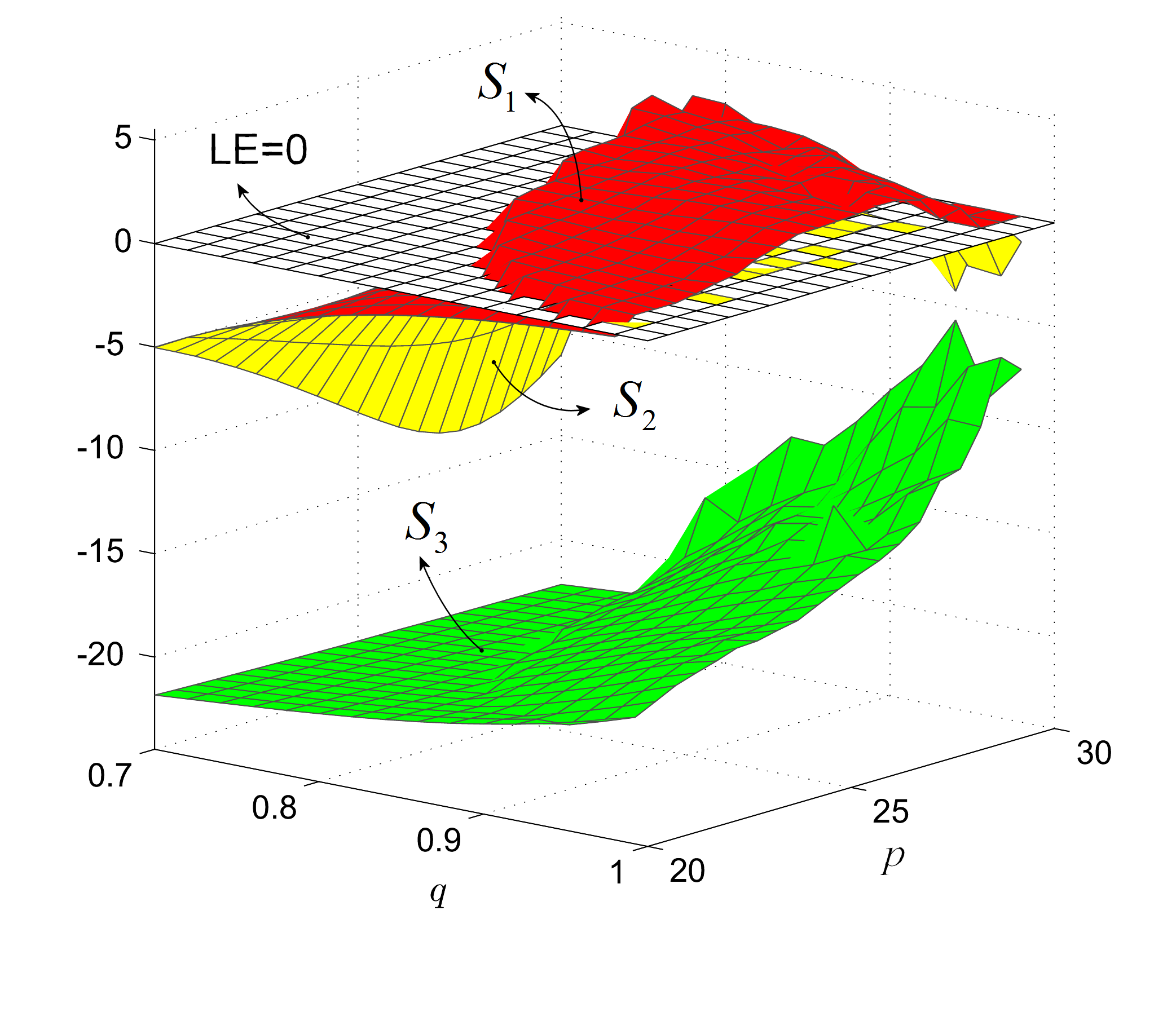}
\caption{LEs of Chen's system of FO represented by function of two variables: $q$ and parameter $p$. Surfaces $S_i$, for $i=1,2,3$, represents $LE(q,p)$.}
\label{fig6}       
\end{center}
\end{figure*}

\clearpage

\section*{Appendices}

\addcontentsline{toc}{section}{Appendices}
\renewcommand{\thesubsection}{\Alph{subsection}}
\subsection{\textup {Program for LEs of FO}}\label{a}
\vspace{-5mm}
\begin{multicols}{2}

\begin{lstlisting}%[multicols=2]%[    basicstyle=\footnotesize, %or \small or \footnotesize etc.]

function [t,LE]=FO_Lyapunov(ne,ext_fcn,t_start,h_norm,t_end,x_start,h,q,out);

%
%    Program to compute LEs of systems of FO.
%
%    The program uses a fast variant of the
%    predictor-corrector Adams-Bashforth-Moulton,
%    "FDE12.m" for FDEs, by Roberto Garrappa:
%
%    https://goo.gl/XScYmi
%
%    m-file required: FDE12 and
%    the function containing the extended system
%    (see e.g. LE_RF.m).
%
%    FO_Lyapunov.m was developed, by
%    modifying the program Lyapunov.m,
%    by V.N. Govorukhin:
%
%    https://goo.gl/wZVCtg
%
%    FO_Lyapunov.m, FDE12.m and LE_RF.m
%    must be in the same folder.
%
%    How to use it:
%      [t,LE]=FO_Lyapunov(ne,ext_fcn,t_start,...
%      h_norm,t_end,x_start,h,q,out);
%
%    Input:
%      ne - system dimension;
%      ext_fcn - function containing the extended
%      system;
%      t_start, t_end - time span;
%      h_norm - step for Gram-Schmidt
%      renormalization;
%      x_start - initial condition;
%      outp - priniting step of LEs;
%      ioutp==0 - no print.
%
%    Output:
%      t - time values;
%      LE Lyapunov exponents to each time value.
%
%    Example of use for the RF system:
%    [t,LE]=FO_Lyapunov(3,@LE_RF,0,0.02,300,...
%    [0.1;0;1;0.1],0.005,0.999,1000);
%
%    The program is presented in:
%
%    Marius-F. Danca and N. Kuznetsov,
%    Matlab code for Lyapunov exponents of
%    fractional order systems
%
%
hold on;

% Memory allocation
x=zeros(ne*(ne+1),1);
x0=x;
c=zeros(ne,1);
gsc=c; zn=c;
n_it = round((t_end-t_start)/h_norm);

% Initial values
    x(1:ne)=x_start;
    i=1;
    while i<=ne
        x((ne+1)*i)=1.0;
        i=i+1;
    end
    t=t_start;
% Main loop
it=1;
while it<=n_it
%       Solution of extended ODE system
        [T,Y] = FDE12(q,ext_fcn,t,t+h_norm,x,h);
        t=t+h_norm;
        Y=transpose(Y);
        x=Y(size(Y,1),:); %solution at t+h_norm
        i=1;
        while i<=ne
            j=1;
            while j<=ne;
                x0(ne*i+j)=x(ne*j+i);
                j=j+1;
            end;
            i=i+1;
        end;
%       orthonormal Gram-Schmidt basis
        zn(1)=0.0;
        j=1;
        while j<=ne
            zn(1)=zn(1)+x0(ne*j+1)*x0(ne*j+1);
            j=j+1;
        end;
        zn(1)=sqrt(zn(1));
        j=1;
        while j<=ne
            x0(ne*j+1)=x0(ne*j+1)/zn(1);
            j=j+1;
        end
        j=2;
        while j<=ne
            k=1;
            while k<=j-1
                gsc(k)=0.0;
                l=1;
                while l<=ne;
                    gsc(k)=gsc(k)+x0(ne*l+j)*x0(ne*l+k);
                    l=l+1;
                end
                k=k+1;
            end
            k=1;
            while k<=ne
                l=1;
                while l<=j-1
                    x0(ne*k+j)=x0(ne*k+j)-gsc(l)*x0(ne*k+l);
                    l=l+1;
                end
                k=k+1;
            end;
            zn(j)=0.0;
            k=1;
            while k<=ne
                zn(j)=zn(j)+x0(ne*k+j)*x0(ne*k+j);
                k=k+1;
            end
            zn(j)=sqrt(zn(j));
            k=1;
            while k<=ne
                x0(ne*k+j)=x0(ne*k+j)/zn(j);
                k=k+1;
            end
            j=j+1;
        end
%       update running vector magnitudes
        k=1;
        while k<=ne;
            c(k)=c(k)+log(zn(k));
            k=k+1;
        end;
%       normalize exponent
        k=1;
        while k<=ne
            LE(k)=c(k)/(t-t_start);
            k=k+1;
        end
        i=1;
        while i<=ne
            j=1;
            while j<=ne;
                x(ne*j+i)=x0(ne*i+j);
                j=j+1;
            end
            i=i+1;
        end;
        x=transpose(x);
        it=it+1;
%       print and plot the results
        if (mod(it,out)==0) %       <@\textcolor{red}{(*)}@>
            fprintf('%10.2f %10.4f %10.4f...
             %10.4f\n',[t,LE]); % <@\textcolor{red}{(*)}@>
        end; %                      <@\textcolor{red}{(*)}@>
        plot(t,LE) %                <@\textcolor{red}{(*)}@>
end
% displays the box outline around axes
xlabel('t','fontsize',16) %         <@\textcolor{red}{(*)}@>
ylabel('LEs','fontsize',14) %       <@\textcolor{red}{(*)}@>
set(gca,'fontsize',14)%             <@\textcolor{red}{(*)}@>
box on;%                            <@\textcolor{red}{(*)}@>
line([0,t],[0,0],'color','k')%      <@\textcolor{red}{(*)}@>


\end{lstlisting}

\subsection{\textup {Function LE\_RF.m}}\label{b}
\begin{lstlisting}[belowskip=0pt]%[    basicstyle=\footnotesize, %or \small or \footnotesize etc.]

function f=LE_RF(t,x)

%Output data must be a column vector
f=zeros(size(x));

%variables allocated to the variational equations
X= [x(4), x(7), x(10);
    x(5), x(8), x(11);
    x(6), x(9), x(12)];

%RF equations
f(1)=x(2)*(x(3)-1+x(1)*x(1))+0.1*x(1);
f(2)=x(1)*(3*x(3)+1-x(1)*x(1))+0.1*x(2);
f(3)=-2*x(3)*(0.98+x(1)*x(2));

%Jacobian matrix
J=[2*x(1)*x(2)+0.1, x(1)*x(1)+x(3)-1, x(2);
     -3*x(1)*x(1)+3*x(3)+1,0.1,3*x(1);
     -2*x(2)*x(3),-2*x(1)*x(3),-2*(x(1)*x(2)+0.98)];

%Righthand side of variational equations
<@\textcolor{red}{f(4:12)=J*X}@>; % To be modified if ne>3

\end{lstlisting}

\subsection{\textup {Program for LEs as function on $p$}}\label{c}
\begin{lstlisting}[belowskip=0pt]%[    basicstyle=\footnotesize, %or \small or \footnotesize e

function run_Lyapunov_p(ne,ext_fcn,t_start,h_norm,t_end,x_start,h,q,p_min,p_max,n);
hold on;
p_step=(p_max-p_min)/n
p=p_min;
while p<=p_max
    LE=FO_Lyapunov_p(ne,ext_fcn,t_start,h_norm,t_end,x_start,h,q,p);
    p=p+p_step
    plot(p,LE);
end
\end{lstlisting}

\subsection{\textup {Function LE\_RF\_p.m}}\label{d}

\begin{lstlisting}[belowskip=0pt]%[    basicstyle=\footnotesize, %or \small or \footnotesize e


function f=LE_RF_p(t,x,p)
%p is the parameter
f=zeros(size(x));
X= [x(4), x(7), x(10);
x(5), x(8), x(11);
x(6), x(9), x(12)];
%RF equations
f(1)=x(2)*(x(3)-1+x(1)*x(1))+0.1*x(1);
f(2)=x(1)*(3*x(3)+1-x(1)*x(1))+0.1*x(2);
f(3)=-2*x(3)*(p+x(1)*x(2));
%Jacobian matrix
J=[2*x(1)*x(2)+0.1, x(1)*x(1)+x(3)-1, x(2);
-3*x(1)*x(1)+3*x(3)+1,0.1,3*x(1);
-2*x(2)*x(3),-2*x(1)*x(3),-2*(x(1)*x(2)+p)];
<@\textcolor{red}{f(4:12)=J*X}@>; % To be modified if ne>3

\end{lstlisting}

\subsection{\textup {Program for LEs as function of $q$}}\label{e}
\begin{lstlisting}[belowskip=0pt]%[    basicstyle=\footnotesize, %or \small or \footnotesize e


function run_FO_Lyapunov_q(ne,ext_fcn,t_start,h_norm,t_end,x_start,h,q_min,q_max,n);
hold on;
q_step=(q_max-q_min)/n;
q=q_min;
while q<q_max
[t,LE]=FO_Lyapunov_q(ne,ext_fcn,t_start,h_norm,t_end,x_start,h,q);
q=q+q_step;
fprintf('q=%10.4f\n %10.4f', q);
plot(q,LE);
end

\end{lstlisting}

\end{multicols}

\end{document}